\documentclass[nofootinbib,twocolumn]{revtex4-2}

\usepackage{graphicx}
\usepackage{dcolumn}
\usepackage{bm}
\usepackage{amsfonts,amsmath,amssymb}
\usepackage[colorlinks,
            linkcolor=red,
            anchorcolor=blue,
            citecolor=green]{hyperref}

\def\be{\begin{equation}}
\def\ee{\end{equation}}
\def\ba{\begin{eqnarray}}
\def\ea{\end{eqnarray}}

\def\no{\nonumber}

\allowdisplaybreaks

\begin{document}

\vspace{-2cm}

\title{Regular black holes and reductions of thermodynamic phase spaces}

\author{Meng-Sen Ma$^{1,2}$}
\email{mengsenma@gmail.com}
\author{Huai-Fan Li$^{2,3}$}
\email{ huaifan999@163.com}
\author{Jian-Hua Shi$^{1,2}$}
\email{dtsjh@139.com}
\affiliation{$^1$Department of Physics, Shanxi Datong
University,  Datong 037009, China
\\
$^2$Institute of Theoretical Physics, Shanxi Datong
University, Datong 037009, China
\\
$^3$ College of General Education, Shanxi College of Technology, Shuozhou 036000, China}

\begin{abstract}

\noindent 
The thermodynamic inconsistency observed in regular black holes is resolved through the framework of reduced thermodynamic phase spaces. We demonstrate that regular black holes are essentially induced from singular black holes by adding an extra requirement, which imposes a constraint among black hole parameters. This constraint reduces the thermodynamic phase space, rendering the standard form of the first law of black hole thermodynamics inapplicable. Accordingly, we propose a novel methodology to study the thermodynamic properties of regular black holes. Thermodynamic quantities must be defined in the full, unconstrained thermodynamic phase space of the underlying singular black holes, only afterward is the constraint imposed to derive the consistent and meaningful thermodynamic quantities of the regular black holes. Crucially, this framework extends beyond regular black holes and applies universally to any black hole with this kind of constraint.

\vspace{-3pt}

\end{abstract}


\maketitle

\section{Introduction} 
Black holes are often accompanied by spacetime singularities. Within these regions, the structure of spacetime breaks down and physical laws cease to apply. Resolving the black hole singularity problem is widely believed to require a full theory of quantum gravity.
Intriguingly, even within the frameworks of classical or semi-classical gravity, black holes without physical singularities can be constructed. These are known as regular or nonsingular black holes. The first such solution was proposed by Bardeen in 1968\cite{Bardeen.87.1968}. Choosing well-behaved energy density, Dymnikova presented an alternative regular black hole\cite{Dymnikova.235.1992}. In a similar way but for different purposes, Hayward proposed the regular black hole that is now named after him\cite{Hayward.031103.2006}. Nicolini and collaborators replaced the conventional Dirac-delta source with a Gaussian matter distribution by incorporating effects from noncommutative geometry and derived the ``noncommutative geometry-inspired black holes", which are also essentially nonsingular\cite{Nicolini.547.2006, Nicolini.1229.2009}. Besides, it has been found that regular black holes can also be obtained by introducing phantom scalar fields as the gravitational source\cite{Bronnikov.251101.2006}.

In \cite{AyonBeato.5056.1998}, Ay{\'o}n-Beato et al. proposed the first regular black hole in general relativity by introducing a nonlinear electrodynamic field as the source. Shortly afterwards, they found that the Bardeen black hole can also be derived as the gravitational field generated by a nonlinear magnetic monopole\cite{AyonBeato.149.2000}. 
After that, many other regular black holes in general relativity and modified gravity were constructed with or without nonlinear electromagnetic sources\cite{Bronnikov.044005.2001, Burinskii.104017.2002, Dymnikova.4417.2004, Breton.643.2005, Berej.885.2006, Balart.124045.2014, Ma.529.2015, Rodrigues.024062.2016, Fan.124027.2016, Nojiri.104008.2017, Ghosh.104050.2018, Maeda.108.2022, Li.104046.2024}, and even rotating regular black holes can also be systematically constructed\cite{Bambi.329.2013, Ghosh.103006.2014, Toshmatov.084037.2017}. Recently, it was discovered that regular black holes can not only have de Sitter cores but also can be present with Minkowski cores\cite{Simpson.8.2020, Ling.075009.2023}. For details, please refer to the recent review article\cite{Lan.202.2023}.

Beyond the issue of spacetime singularities, the thermodynamic properties of black holes constitute another major focus of modern gravitational physics. Since the discovery of Hawking radiation and Bekenstein's proposal for black hole entropy, black holes have been recognized as thermodynamic systems possessing temperature, entropy, and other thermodynamic characteristics, which are governed by the four laws of black hole thermodynamics. However, in \cite{Ma.245014.2014}, we found that, for many regular black holes, there is an inconsistency between the temperature and the entropy. If one insists that black hole entropy strictly adheres to the area law, the temperature $\left.\frac{\partial M}{\partial S}\right|_{...}$ derived from the first law is not equal to the black hole temperature. Conversely, if one substitutes the black hole temperature into the first law and integrates $\int\left.\frac{dM}{T}\right|_{...}$, the derived entropy frequently deviates from the area law. In previous studies on the thermodynamics of regular black holes, people had to choose one and abandon the other\cite{Myung.012.2007, Banerjee.124035.2008, Nicolini.097.2011, Akbar.070401.2012, Smailagic.1350010.2013, Zhang.145007.2018, Li.1950336.2019, Tzikas.219.2019, Guo.025402.2022,Simovic.044029.2024, Bakopoulos.L101502.2024}. In  \cite{Ma.245014.2014}, we attributed this inconsistency to the occurrence of black hole mass $M$ in the  Lagrangian of matter fields and obtained a corrected first law of black hole thermodynamics.

Now we find that our previous explanation does not delve into the essence of the problem. The root of this inconsistency lies in the fact that in the process of constructing regular black holes through matter fields, one will inevitably introduce a constraint among black hole parameters. This constraint not only reduces the dimension of the thermodynamic phase space, but also renders the thermodynamic quantities not independent of each other and modifies the first law of thermodynamics.
Consequently, the temperature derived geometrically (e.g., via surface gravity) no longer automatically satisfies the first law when combined with the Bekenstein-Hawking entropy. 

Regular black holes are essentially obtained by imposing additional constraint relations on underlying singular black holes. Thermodynamic inconsistency is inherent in these regular black holes. Consequently, to fully understand their thermodynamic properties, we must go beyond the reduced thermodynamic phase space intrinsic to the regular black holes themselves. Instead, our analysis should focus on the full thermodynamic phase space of the underlying singular black holes. Specifically, thermodynamic quantities should first be defined within the singular black holes. Only after this should the constraints be applied to derive the consistent thermodynamic properties of the regular black holes. 
From this novel perspective, this work aims to systematically analyze thermodynamic consistency issues in regular black holes and discuss the deformation of thermodynamic structures induced by additional constraints.

\section{Reductions of thermodynamic phase spaces}
Thermodynamic phase space is a $(2n+1)$-dimensional manifold $\mathcal{T}$ with coordinates $(\Phi, E^a, I^a), a=1,2,...n$, where $E^a$ and $I^a$ are extensive and intensive parameters, respectively, and are conjugate to each other. $\Phi=\Phi(E^a)$ is the thermodynamic potential\cite{Quevedo.013506.2007}. After introducing the Gibbs 1-form $\Theta=d\Phi-\delta_{ab}I^adE^b$, we obtain the contact manifold, $(\mathcal{T},\Theta)$, with contact structure. The space of thermodynamic equilibrium states $\mathcal{E}$ is $(\Phi(E^a), E^a, I^a(E^a))$, the $n$-dimensional subspace of $\mathcal{T}$ with the defining parameters $(E^a)$ as coordinates. 

Below we will take the Kerr-Newman-AdS(KN-AdS) black hole as an example to demonstrate how the reductions of thermodynamic phase spaces affect the thermodynamic quantities of black holes. As is shown in\cite{Caldarelli.399.2000}, there is an relation between the black hole parameters $(M,J,Q,S, \Lambda)$,
\be\label{KNAdS_M}
M^2=J^2 \left(\frac{\pi }{S}+\frac{\Lambda }{3}\right)+\frac{S }{4 \pi }\left(\frac{\pi  Q^2}{S}+\frac{\Lambda  S}{3 \pi }+1\right)^2.
\ee
In the parameter space $\{S,Q,J,P\}$, where the cosmological constant is considered variable and identified with the thermodynamic pressure $P=-\frac{\Lambda}{8\pi}$\cite{Kastor.195011.2009, Dolan.235017.2011, Cvetic:2011, Kubiznak.033.2012}, we have the first law and the Smarr formula
\ba
dM&=&TdS+\Psi dQ+\Omega dJ +VdP, \\
M&=&2TS+2\Omega J+\Psi Q-2PV,
\ea
where the temperature is
\be\label{KNAdS_T}
T=\left.\frac{\partial M}{\partial S}\right|_{Q,J,P}.
\ee
Black hole temperature is a kinematical effect, which can be derived from many other approaches, such as the elimination of the conical singularity and the tunneling method. We call the black hole temperature derived by these methods the geometric temperature, denoted as $T_g$. For comparison, the temperature in Eq.(\ref{KNAdS_T}), obtained through the first law of thermodynamics, is called the thermodynamic temperature. It can be easily checked that the two temperatures are equal for the KN-AdS black hole.

Now we set $(J=0,~\Lambda=0)$, which corresponds to the RN black hole. After adding the two constraints, the parameter space is reduced to $\{S,Q\}$ and the first law becomes
\be\label{RN_1st}
dM=TdS+\Psi dQ.
\ee
The mass relation becomes
\be\label{RN_M}
M^2=\frac{S}{4\pi}\left(\frac{\pi  Q^2}{S}+1\right)^2.
\ee
The metric function of RN black hole is 
\be
f(r)=1-\frac{2M}{r}+\frac{Q^2}{r^2}.
\ee
Now we can derive the temperature of the RN black hole in three ways.
The first way is to calculate the geometric temperature according to the metric function,
\be\label{RN_T}
T_g=\frac{\kappa}{2\pi}=\frac{1}{4\pi r_{+}}\left(1-\frac{Q^2}{r_{+}^2}\right).
\ee
Besides, we have two ways to derive the thermodynamic temperature. One can set $(J=0,~\Lambda=0)$ in Eq.(\ref{KNAdS_T}) directly and obtain $T^{(I)}$. Or one can recalculate the thermodynamic temperature on the basis of Eq.(\ref{RN_1st}) and Eq.(\ref{RN_M}) and obtain $T^{(II)}$. It can be found that the three ways give the same result, namely
\ba\label{RNAdS}
M(S,Q,J,P) \quad \xrightarrow[J=0]{~~~\Lambda=0~~~} \quad M(S,Q) \qquad \quad \no \\
\downarrow  \qquad\qquad\qquad\qquad\qquad\qquad \downarrow  \qquad \qquad  \\
\left.\frac{\partial M}{\partial S}\right|_{Q,J,P} \xrightarrow[J=0]{~\Lambda=0~} T^{(I)} =T_g=T^{(II)}\equiv\left.\frac{\partial M}{\partial S}\right|_{Q}.\no
\ea

Particular attention must be paid to the two thermodynamic temperatures. The former calculates the temperature in the full parameter space and then restricts it with constraint conditions. The latter is to consider the constraint conditions first to obtain the reduced parameter space, and then calculate the temperature in this reduced space.

Next, further set $M=Q$ based on the RN black hole to obtain the extremal RN black hole, inconsistencies begin to emerge. For the extremal RN black hole, the inner horizon and the outer event horizon coincide, and $r_{+}=Q$. Clearly, according to Eq.(\ref{RN_T}), the geometric temperature is zero\footnote{The temperature of extremal black holes is still controversial. As pointed out in \cite{Hawking:1994}, no conical singularity exists in the extremal case and the temperature can be arbitrary.}. One can see  that $T^{(I)}$ is also zero, but $T^{(II)}$ is nonzero. After considering the constraint $M=Q$, we are left with only one free parameter, $S$. Now the mass function is
\be
M=\sqrt{S/\pi},
\ee
from which and the first law one can derive 
\be
T^{(II)}=\frac{dM}{dS}=\frac{1}{2\pi r_{+}} \neq 0,
\ee
namely
\ba\label{exRN}
M(S,Q) \qquad \xrightarrow{~~~~M=Q~~~~}  \qquad M(S) \qquad\nonumber \\
\downarrow  \qquad\qquad\qquad\qquad\qquad\qquad  \downarrow  \quad\qquad   \\
\left.\frac{\partial M}{\partial S}\right|_{Q}  \xrightarrow{M=Q~~} T^{(I)}=T_g \neq T^{(II)}\equiv\frac{d M}{d S}.\nonumber
\ea
The root of this inconsistency is that the constraint $M=Q$ modifies the first law of thermodynamics. We first differentiate the constraint to obtain $dM=dQ$, then substitute it into Eq.(\ref{RN_1st}) to eliminate $dQ$, we get
\be
dM=\frac{T^{(I)}}{1-\Psi}dS=T^{(II)}dS,
\ee
which gives the relation between the $T^{(I)}$ and $T^{(II)}$. The electric potential is $\Psi=Q/r_{+}$. In the extremal case, both $T^{(I)}$ and $1-\Psi$ tend to zero, but their ratio gives a nonzero result,
\be
\frac{T^{(I)}}{1-\Psi}=\frac{r_{+}+Q}{4\pi r_{+}^2}=\frac{1}{2\pi r_{+}}=T^{(II)}.
\ee
Therefore, $T^{(II)}$ is essentially an effective temperature and has no physical meaning in itself.

By comparing Eq.(\ref{RNAdS}) and Eq.(\ref{exRN}), it can be seen that the constraints $J=0$ and $\Lambda=0$ merely reduces the dimension of the thermodynamic phase space and does not modify the first law of thermodynamics. Therefore, the temperatures $T^{(I)}$ and $T^{(II)}$ are equal. While the kind of constraint involving $M$, such as $M=Q$, not only reduces the phase space but also modifies the first law of thermodynamics, which leads to the inconsistency of the two temperatures.

\vspace{2pt}

\section{Temperatures of regular black holes}
The example of the extremal RN black hole is just an appetizer, but it provides inspiration for solving the thermodynamic inconsistency of regular black holes, because there is an implicit extra constraint in regular black holes.

% Black holes, as thermodynamic systems, should satisfy the first law of thermodynamics,
% \be\label{1st}
% \delta M=T\delta S+\Omega\delta J +...,
% \ee
% where the ``..." denote the possible additional contributions from long-range fields and all dimensionful parameters\cite{Wald:2000}.

We are concerned only with static spherically symmetric black holes in general relativity with matter fields. The line element can be written as
\be\label{staticmetric}
ds^2=-f(r)dt^2+f(r)^{-1}dr^2 + r^2d\Omega^2.
\ee
In the absence of quantum corrections and higher curvature corrections, there is no doubt that the black hole entropy should satisfy the Bekenstein-Hawking area law, 
\be
S=\frac{A}{4}=\pi r_{+}^2,
\ee
which can be verified through many approaches\cite{Wald:1993, Jacobson:1994, Wald:2001}.

The geometric temperature is given by
\be\label{Hawking_T}
T_g=\frac{f'(r_{+})}{4\pi}.
\ee
In \cite{Ma.245014.2014}, we found that, for many regular black holes, $T_g$ does not satisfy the first law of black hole thermodynamics. In other words, $T_g \neq \left.\frac{\partial M}{\partial S}\right|_{...}$. Next, we will present the essence of the issue.

For simplicity, we take the metric function in the following form
\be\label{efm}
f(r)=1-\frac{2m(r)}{r}.
\ee
According to the Einstein field equations, 
\be
G_{\mu\nu}=8\pi T_{\mu\nu},
\ee
$m(r)$ should satisfy
\be\label{mr}
\frac{d m(r)}{d r}=-4\pi r^2 T^0_{~0}.
\ee

After integration, one can obtain
\be\label{mr2}
m(r)=M+4\pi\int_{r}^{\infty}r^2T^0_{~0}dr,
\ee
where $M$ is an integration constant, which is taken to fulfill the boundary condition $M=\lim\limits_{r\rightarrow \infty}m(r)$. Obviously, $M$ is just the ADM mass in the asymptotically flat spacetime.

Inspired by the work\cite{Fan.124027.2016}, to obtain a regular black hole, we must add another condition 
\be\label{m_constraint}
\lim_{r\rightarrow 0}m(r)=0.
\ee
It is a sufficient but not necessary condition to judge the regularity of a black hole solution.

Combing Eq.(\ref{mr2}) and Eq.(\ref{m_constraint}), we have
\be\label{M_constraint}
M=-4\pi\int_{0}^{\infty}r^2T^0_{~0}dr.
\ee
Generally, $T^0_{~0}=T^0_{~0}(r,...)$, where `..." represents other parameters from the matter fields. After integration, we are left with the relation $M=M(...)$, 
which imposes an additional constraint between black hole parameters, leading to a reduction of the thermodynamic phase space. It is the chief culprit that leads to the inconsistency of thermodynamic quantities.

For clarity, let us take the Bardeen regular black hole as an example. For simplicity, here we adopt a slightly different Lagrangian from \cite{AyonBeato.149.2000, Fan.124027.2016}
\be\label{Bardeen_L}
\mathcal{L}=-\alpha  \left(\frac{\sqrt{F/2}}{1+\sqrt{F/2}}\right)^{5/2},
\ee
which includes only one parameter $\alpha$ that is initially independent of the black hole mass and the magnetic charge. $\alpha$ is a positive coupling constant with the dimension of $[L]^{-2}$ and $F=F_{\mu\nu}F^{\mu\nu}$. 

In the spherically symmetric case with pure magnetic fields, $F=\frac{2Q^2}{r^4}$, where $Q$ is the magnetic charge, an integration constant.

The energy-momentum tensor can be derived according to
\be
T_{\alpha\beta}=g_{\alpha\beta}\mathcal{L}+4\mathcal{L}_FF_{\alpha\mu}F^{\mu}_{~\beta},
\ee
where $\mathcal{L}_F\equiv \partial{\mathcal{L}}/\partial{F}$. In the spherically symmetric case with pure magnetic charge, we have $T^0_{~0}=\mathcal{L}$. Substituting it into Eq.(\ref{mr2}) and integrating, we obtain
\ba\label{Bardeen}
m(r)&=&M+4\pi\int_{r}^{\infty}r^2\mathcal{L}dr \nonumber \\
&=&M-\frac{4\pi}{3} \alpha  Q^{3/2}\left[1-\frac{r^3}{\left(Q+r^2\right)^{3/2}}\right].
\ea
It is a usual singular black hole. From $f(r_{+})=0$, in the parameter space $\{S,Q,\alpha\}$, we can express $M$ as
\be\label{M_without_cons}
M=\frac{\sqrt{S}}{2 \sqrt{\pi }}+\frac{4\pi}{3} \alpha  Q^{3/2}\left[1-\frac{S^{3/2}}{(\pi  Q+S)^{3/2}}\right],
\ee
from which we have the first law of black hole thermodynamics and the Smarr formula
\ba
dM&=&TdS+\phi dQ+\mathcal{A}d\alpha, \label{first_law} \\
M&=&2TS+2\phi Q-2\mathcal{A}\alpha, \label{smarr}
\ea
where
\be
T=\left.\frac{\partial M}{\partial S}\right|_{Q,\alpha}, \quad \phi=\left.\frac{\partial M}{\partial Q}\right|_{S,\alpha}, \quad \mathcal{A}=\left.\frac{\partial M}{\partial \alpha}\right|_{S,Q}.
\ee
Because there is no constraint among the black hole parameters in this case, one can easily check that the temperature derived from the first law is indeed equal to the geometric temperature derived according to Eq.(\ref{Hawking_T}),
\be\label{singular_T}
T=\frac{1}{4 \pi  r_{+}}-2 \alpha  r_{+} \left(\frac{Q}{Q+r_{+}^2}\right)^{5/2}.
\ee

After considering Eq.(\ref{m_constraint}), we obtain from Eq.(\ref{Bardeen}) 
\be\label{Bardeen_cons}
M=\frac{4\pi}{3} \alpha  Q^{3/2},
\ee
which imposes an extra restriction on the parameter space 
$\{S,Q,\alpha\}$ and plays a role similar to the extremal condition of RN black hole. 

Now the metric function is 
\be
f(r)=1-\frac{8\pi}{3} \frac{\alpha  Q^{3/2}r^2}{\left(Q+r^2\right)^{3/2}}=1-\frac{2 M r^2}{\left(Q+r^2\right)^{3/2}},
\ee
which is just the Bardeen black hole. Substituting this metric function into Eq.(\ref{Hawking_T}) or substituting the constraint in Eq.(\ref{Bardeen_cons}) into Eq.(\ref{singular_T}), we have
\be\label{Bardeen_T1}
T_g=\frac{r_{+}^2-2Q}{4\pi r_{+}(Q+r_{+}^2)}=T^{(I)}.
\ee

Combing Eq.(\ref{M_without_cons}) and Eq.(\ref{Bardeen_cons}), one can eliminate $\alpha$ or $Q$ to obtain the reduced thermodynamic phase space $\{S, Q\}$ or $\{S, \alpha\}$, respectively. For simplicity, we consider the $\{S, Q\}$ space. Now $M$ can be expressed as
\be
M=\frac{(\pi  Q+S)^{3/2}}{2 \sqrt{\pi } S},
\ee
from which one can derive the thermodynamic temperature
\be\label{Bardeen_T2}
T^{(II)}=\left.\frac{\partial M}{\partial S}\right|_{Q}=\frac{\left(r_+^2-2 Q\right) \sqrt{Q+r_+^2}}{4 \pi  r_+^4}.
\ee
Obviously, we have $T^{(I)}\neq T^{(II)}$.

To clarify this issue, we take the total differential of Eq.(\ref{Bardeen_cons}), 
\be
dM=\frac{4\pi}{3}  Q^{3/2}d\alpha +2\pi \alpha Q^{1/2}dQ,
\ee
and then substitute it into Eq.(\ref{first_law}) to eliminate $d\alpha$. In the reduced thermodynamic phase space $\{S,Q\}$, we have
\ba\label{1stlawQ2}
dM=\frac{T^{(I)}}{1-\frac{3\mathcal{A}}{4\pi Q^{3/2}}}dS+\frac{\phi-\frac{3\alpha \mathcal{A}}{2Q}}{1-\frac{3\mathcal{A}}{4\pi Q^{3/2}}}dQ
\ea
with 
\be
1-\frac{3\mathcal{A}}{4\pi Q^{3/2}}=\frac{r^3}{\left(Q+r^2\right)^{3/2}}.
\ee

By comparing with Eq.(\ref{Bardeen_T2}), it can be found that 
\be
T^{(II)}=\frac{T^{(I)}}{1-\frac{3\mathcal{A}}{4\pi Q^{3/2}}}.
\ee
We now understand that the inconsistency between the temperature of regular black holes and the first Law of thermodynamics originates from the introduction of an additional constraint. This constraint not only reduces the dimensionality of the thermodynamic phase space but also modifies the first law in the reduced phase space. 

In \cite{Ma.245014.2014}, we also discussed other two types of regular black holes -- the noncommutative geometry inspired Schwarzschild black hole(NCSBH)\cite{Nicolini.547.2006} and Dymnikova's nonsingular black hole\cite{Dymnikova.235.1992}. These solutions are derived from some strange gravitational sources, respectively 
\ba
\rho&=&\frac{M}{(4 \pi \theta)^{3 / 2}} \exp \left(-r^2 / 4 \theta\right), \\
\rho&=&\varepsilon_0 \exp \left(-\frac{r^3}{r_0^2 r_g}\right),
\ea
which has no corresponding Lagrangian. The thermodynamic inconsistency in the two models also stems from the reason analyzed above. In fact, after substituting the constraints Eq.(\ref{Bardeen_cons}) into the nonlinear electrodynamic Lagrangian, Eq.(\ref{Bardeen_L}), we can get
\be
\rho=-T^0_{~0}=\frac{3 M}{4 \pi  Q^{3/2}}\left(\frac{Q}{Q+r^2}\right)^{5/2}.
\ee

Furthermore, to our knowledge, there do exist regular black hole models free from such inconsistencies. In \cite{Ma:2017}, we constructed several regular black holes in (2+1)-dimensional spacetime, the thermodynamic quantities of which indeed satisfy the first law. This is because there are naturally no additional constraints in the (2+1)-dimensional model. The field equation has the form
\be
\frac{f^{\prime}(r)}{2 r}+\Lambda=8 \pi\left[L(r)-\frac{4 q}{r} E(r)\right],
\ee
where $L(r)$ and $E(r)$ are the Lagrangian of the nonlinear electromagnetic source and electric field, respectively. The metric function takes the form
\be
f(r)=-M+\frac{r^2}{l^2}+k(r).
\ee
To construct a regular black hole, we should add the constraint
\be
\lim _{r \rightarrow 0} \frac{f^{\prime}(r)}{r}=\text { constant }.
\ee
Clearly, after differentiation, the parameter $M$ disappears in the constraint equation, leaving behind an identity.

In \cite{Bueno.139260.2025}, some regular black holes were constructed from pure gravity, which is a higher-curvature theory of gravity, but not GR. The conditions for singularity resolution do not introduce any constraints on the ADM mass $M$ and other parameters. This guarantees that the first law of thermodynamics still holds for these regular black holes.

\vspace{2pt}

\section{Thermodynamics of regular black holes}
In this part, we will discuss how to define other thermodynamic quantities of regular black holes, such as heat capacity and free energy. 

For comparison, we first consider the RN black hole. As discussed above, one can directly define and calculate thermodynamic quantities in the thermodynamic phase space of RN black hole itself. Besides, one can also calculate the thermodynamic quantities in the phase space of the RN-AdS black hole, and then set $\Lambda=0$. The results obtained by the two ways are the same. For example, the heat capacity at constant electric charge can be derived in two ways,
\be
C_{Q,\Lambda} \equiv \left.\frac{\partial M}{\partial T}\right|_{Q,\Lambda} \xrightarrow{~~\Lambda=0~~} C_Q \equiv \left.\frac{\partial M}{\partial T}\right|_{Q}.
\ee

For the Bardeen black hole, the thermodynamic parameter space is $(S, Q)$, but we cannot define thermodynamic quantities in this space directly. According to the previous discussion, we are certain that the thermodynamic temperature $T^{(II)}$ is not a physical quantity. Thus, we cannot use the non-physical temperature to define the heat capacity. $T^{(I)}=T_g$ is the physical temperature. However, it is inconsistent with the first law of black hole thermodynamics in the parameter space $(S, Q)$. One cannot define the heat capacity in this way, $C_Q=\left.\frac{\partial M}{~~\partial T^{(I)}}\right|_{Q}$. Because the heat capacity is defined as the amount of heat to be supplied to an object to produce a unit change in its temperature. Only when the first law of thermodynamics holds is the absorbed heat equal to $dM$.

To ensure that both the temperature and the first law of thermodynamics are meaningful, the only way is to move beyond the reduced thermodynamic phase space intrinsic to the regular black hole. Before imposing the constraint, the black hole temperature and the first law in the singular black hole are consistent. Therefore, we should focus on the higher-dimensional parameter space $(S, Q, \alpha)$ and define thermodynamic quantities in the singular black hole. Then we impose the constraint to obtain the consistent thermodynamic properties of the regular black holes. 

In our example, combing Eq.(\ref{M_without_cons}) and Eq.(\ref{Bardeen_cons}) to eliminate $M$, we get
\be\label{Bardeen_cons2}
\alpha=\frac{3 }{8 \pi  r_+^2}\left(\frac{Q+r_+^2}{Q}\right){}^{3/2}.
\ee
The correct heat capacity of the Bardeen black hole should be derived in the following way,
\be
C_{Q,\alpha}=\left.\frac{\partial M}{\partial T}\right|_{Q,\alpha} \xrightarrow{(\ref{Bardeen_cons2})} C=\frac{2 \pi  r_+^2 \left(2 Q-r_+^2\right) \left(Q+r_+^2\right)}{4 Q^2-10 Q r_+^2+r_+^4}.
\ee

\begin{figure}[h!]
	\centering{
	  \includegraphics[width=7cm]{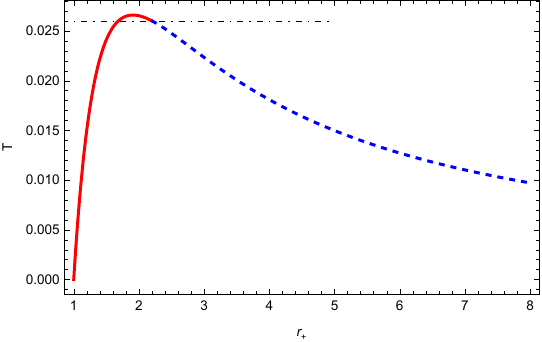}
        \includegraphics[width=7cm]{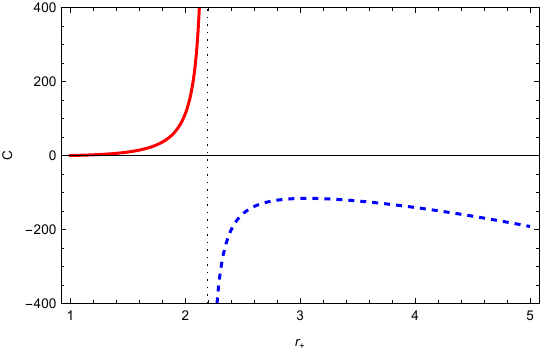}
        \includegraphics[width=7cm]{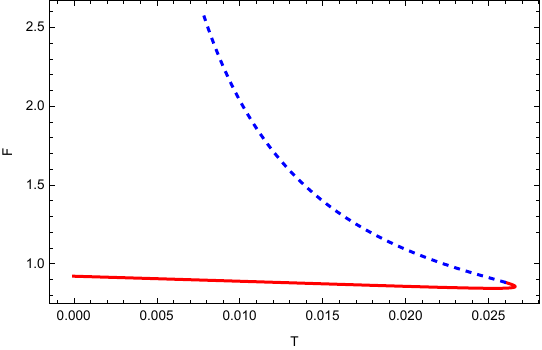}
	\caption{The behaviors of the temperature, the heat capacity and the free energy for the Bardeen black hole. We set $Q=0.5$. The maximum of the temperature lies at $r_{+}=1.907$ and the divergent point of the heat capacity lies at $r_{+}=2.189$. The dot-dashed line in the temperature curve corresponds to $T=0.026$.} \label{figTCF}
	}
\end{figure}

Free energy is defined through the Legendre transformation based on the first law of thermodynamics. We should first define the Helmholtz free energy in the parameter space $(S, Q, \alpha)$ of the singular black hole, and then impose the constraint to get the corresponding free energy of the Bardeen black hole.
\be
F=M-TS \xrightarrow{(\ref{Bardeen_cons2})} F_{\text{Bardeen}}.
\ee
It should be noted that the free energy derived in this way corresponds to the fixed $(Q,\alpha)$ ensemble of the singular black hole. One can also define free energies in other ensembles through the Legendre transformation, then reduce to the free energy of the Bardeen black hole. 

In Fig.\ref{figTCF}, it is shown that the thermodynamic behavior of the Bardeen black hole is similar to that of the RN black hole. However, there is still a major difference. The divergent point of the heat capacity does not correspond to the extremal point of the temperature. This correspondence, however, always holds for singular black holes since they satisfy the first law of thermodynamics, where the heat capacity satisfied $C=\frac{\partial M}{\partial T}=\frac{\partial M/\partial r_{+}}{\partial T/\partial r_{+}}$. For the Bardeen black hole, when $T<0.026$, only the smaller black hole has positive heat capacity. When $T>0.026$, the Bardeen black hole admits two locally stable states with identical temperature: a larger black hole phase and a smaller black hole phase. Furthermore, the $F-T$ curve reveals that the smaller black hole phase possesses lower free energy. Consequently, the smaller black hole state is globally thermodynamically stable.

The thermodynamics of regular black holes can also be investigated in asymptotically AdS spacetime, where we expect the possibility of $P-V$ criticality and a richer phase structure.

In \cite{Huang:2024}, the authors also noticed that the presence of additional constraints renders the thermodynamic quantities of regular black holes no longer independent. They employed the restricted phase space (RPS) formalism and adopted a similar strategy by imposing constraints on the thermodynamic quantities of singular black holes to obtain those of regular black holes, and then used this framework to investigate the thermodynamic properties of several regular black holes.

\vspace{2pt}

\section{Final remarks} 
In this paper, we provide a coherent explanation for the apparent thermodynamic inconsistencies observed in regular black holes by introducing the concept of a reduced thermodynamic phase space. Using the Bardeen black hole as a representative example, we demonstrate that the construction of a regular black hole intrinsically introduces a constraint condition among the black hole parameters. This constraint effectively confines the regular black hole to a reduced thermodynamic phase space, leading to a situation where the standard form of the first law of black hole thermodynamics no longer holds. This resolves the discrepancy between the temperature calculated via geometric methods and that derived from the standard thermodynamic first law.

Building on this observation, we argue that, to properly analyze the thermodynamics of regular black holes, one should not work solely within the reduced phase space of the regular solution. Instead, one must commence from the full thermodynamic phase space of the corresponding singular black hole, compute the relevant thermodynamic quantities, and then impose the constraint to obtain the thermodynamics of the regular black hole. This approach ensures the consistency of the black hole temperature with the first law and yields physically meaningful results for other thermodynamic quantities such as heat capacity and free energy.

Furthermore, the relationship between reduced phase spaces and black hole thermodynamics is not unique to regular black holes. Any black holes with additional constraints among black hole parameters inherently involve a reduction of the thermodynamic phase space. Consequently, the framework developed in this work must be employed to rigorously investigate the thermodynamic properties of such black holes.

\vspace{2pt}

\noindent \textbf{Acknowledgements.} We are pleased to thank Prof. Ren Zhao and Prof. Chao-Guang Huang for helpful discussions. This work is supported in part by Shanxi Provincial Natural Science Foundation of China (Grant No. 202203021221211), the Natural Science Foundation of China (Grant No. 12075143), and Scientific and Technological Innovation Programs of Higher Education Institutions in Shanxi (Grant No. 2021L386).

\noindent \textbf{Conflict of interest} The authors declare that they have no conflict of interest. 

% \vspace{-15.1pt}

\bibliography{Regularblackholes}

\end{document}